\documentclass[prl,floatfix,twocolumn,showpacs]{revtex4}
\usepackage{graphicx}
\begin{document}
\title{Baryons and Pentaquarks in terms of Mesons}
\author{Ramesh Anishetty}
\email[email(corresponding author):]{ramesha@imsc.res.in}
\author{Santosh Kumar Kudtarkar}
\email[email:]{sant@imsc.res.in}
\affiliation{The Institute of Mathematical Sciences, CIT-Campus, Taramani,
Chennai, India}
\date{\today}
\begin{abstract}
We investigate light meson bound state contributions to six quark Green's functions and establish 
that the latter has poles corresponding to  baryon bound states. We estimate  light baryon 
masses including the Roper resonance. Constituent quark model and 
the Parton 
model are natural consequences of this quantum field theoretic picture. We elaborate this analysis to 
pentaquarks and 
heptaquarks. An almost model independent prediction for the mass of $\theta$ pentaquark is 
$1.57Gev$.
\end{abstract}

\maketitle
\begin{figure}
\includegraphics[height=3cm]{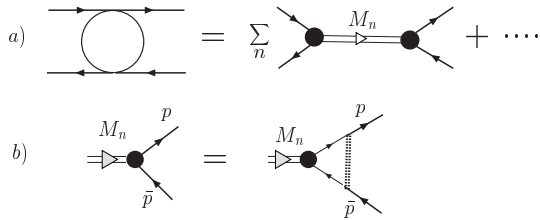}
\caption{(a)The Meson contributions to the quark anti-quark scattering kernel. (b) The 
Bethe-Salpeter equation for the Meson}
\label{fig:scat}
\end{figure}

We look for a baryon pole in six quark Green's functions $\langle qqq \bar q \bar q \bar q \rangle$ 
in 
QCD.
Non-perturbatively this has many contributions, in particular let us concentrate on 
contributions due to quark-quark 
scattering. In a relativistic Quantum field theory like QCD this can be inferred from 
quark-antiquark scattering by crossing symmetry. We assume the color singlet $s$ channel of 
$q \bar q$ scattering does have color singlet meson poles which are the observed pion, $\rho,\omega,K$ 
etc(Fig[\ref{fig:scat}]). The wave-function $\phi_n$ of these mesons are realised as eigensolutions 
of the 
Bethe-Salpeter($BS$) equation wherein a kernel is subsumed to exist from the underlying theory, QCD.
\begin{equation}
\langle q \bar q ~ q \bar q \rangle = \sum_n \phi_n^* \frac{1}{s-M_n^2 +i \epsilon} \phi_n + \cdots
\end{equation}
where $\phi_n$ obeys the $BS$ equation (Fig[\ref{fig:scat}b]). $M_n$ is the mass of the 
meson of 
type $n$. 
\begin{equation}
\phi_n =S(p)~\bar S(\bar p) \int K \phi_n
\label{BS}
\end{equation}
 $S(p)$ and $\bar S(\bar p)$ are the exact propagators for the quarks.
The $BS$ wave-function does
have a spectral representation which follows from causality\cite{deser,meson}.  
It is given by
\begin{equation}
\phi_n=\!\int_0^1\!dy\!\int_0^\infty\!\frac{d\beta~~~ \tilde \phi_n(y,\beta)} 
{yp^2\!+\!(1-y)\bar p^2\!-\!y(1-y)M_n^2\!-\!\delta_n^2\!-\!\beta\!+\!i\epsilon}
\label{BS_spec}
\end{equation}
The spin angular momentum,flavour details are suppressed in the above equation. More precise 
definitions along 
with 
spin-orbit coupling details are presented in \cite{meson} along with asymptotic behaviour of the
$BS$ wave-function which are  
deduced
from the structure of the $BS$ equation.  For completeness we state that there is a generic
spectral representation for the quark propagator satisfying the Schwinger-Dyson($SD$) equation in QCD
\begin{equation}
S(p)= \int_0^\infty d\alpha \frac{\tilde S(\alpha)}{ p^2-\tilde{m}^2-\alpha +i \epsilon}
\label{quark_spec}
\end{equation}
 We have introduced two important quantities\cite{quark} $\tilde{m}$ and $\delta_n$. $\tilde{m}$ is 
the 
threshold 
mass
$i.e.,$ only for momenta $p^2 \geq \tilde m^2$ will the quark propagator have an imaginary part. It 
may in general be a pole if the quark is not confined or a branch cut if it is confined. 
Similarly $\delta_n$ is the relativistic analogue of the size(in units of mass) of the 
meson\cite{meson} of mass $M_n$. In 
particular the 
meson wave-function $\phi_n$ is a function of two complex variables, $p^2$ and $\bar p^2$. 
Only for momenta $p^2 \geq \delta_n^2$ or $\bar p^2 \geq \delta_n^2$ 
the wave-function  $\phi_n$ has singularities. The threshold mass $\tilde m$ is in 
principle determined form the $SD$ equation while the meson size $\delta_n$ and meson mass $M_n$ 
are self-consistently determined from the $BS$ equations. From eq.(\ref{BS}) it is easy to see
that 
\begin{equation}
0\leq \delta_n \leq min (\tilde m , \bar{\tilde m})
\label{bound}
\end{equation}
The lower inequality follows from the the fact that all amplitudes have no singularities in the 
space-like domain(here it amounts to saying the meson has a finite size) while the upper bound
is self-evident from the $BS$ equation, eq.(\ref{BS}) ($\tilde m$ and $\bar{\tilde m}$ being the 
threshold 
masses of the quark and the anti-quark). A more precise estimate of $\delta_n$ demands the 
knowledge of the exact structure of the kernel in the $BS$ equation.  From the definition of 
$\delta_n$(eq.(\ref{BS_spec})), we note that a bound quark in a meson has a threshold singularity
at $\delta_n$ which is less than that of the unbound quark(eq.(\ref{bound})).

\begin{figure}
\includegraphics[width=4cm]{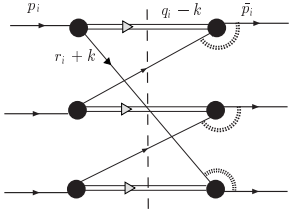}
\caption{Baryon pole(constituent quark picture). }
\label{fig:proton}
\end{figure}

Consider the amplitude of three quarks going to three quarks. We notice that there are contributions
due to three different quark-quark scattering. One such contribution of interest is the one shown in 
Fig[\ref{fig:proton}]. 
This involves one internal loop momentum $k$ as explicitly shown in the Fig[\ref{fig:proton}]. By 
choice we take
$q_i$ and $r_i$(with $r_1=0$) as independent momenta while the external momenta $p_i$ and $\bar p_i$ 
follow from momentum conservation at the vertices.
 To avoid the ambiguity of double counting the quark propagator, we introduce 
the 
kernel used in the  $BS$ equation. Using  spectral representation for  the 
$BS$ wavefunction(eq.(\ref{BS_spec})) this diagram is completely well defined and the loop 
momentum,$k$, 
integration can be done.
 We are interested in the singularity structure of this graph. We apply the Hadamard/Landau 
conditions\cite{eden} on all variables being integrated over, which includes  $y_i$ and $\beta_i$ 
coming from
each of the spectral representations of the meson vertices. To make the computation transparent 
first
notice that we have to consider terms like 
\begin{equation}
\Delta = y_i(k+r_i)^2 -(1-y_i)p_i^2 -y_i(1-y_i)M_{i}^2 -\delta_i^2-\beta_i
\end{equation}
This is equivalent to an effective mass term for the quark propagator which has momentum $(k+r_i)$.
Landau equations demand that $\beta_i=0$ and $y_i=0,1$ or $\frac{\partial \Delta}{\partial y_i}=0$.  
Furthermore looking for the lowest effective mass with respect to $p_i^2$ $i.e.,$ 
$\frac{\partial \Delta}{\partial p_i^2}=0$ we require  $y_i=1$. Consequently, we find $\Delta \approx 
(k+r_i)^2-\delta_i^2$.  To evaluate the nature of the singularity it is convenient to employ 
Cutkosky
cuts\cite{dia} as shown by the dashed line in Fig[\ref{fig:proton}]. With  suitable conditions on 
$\tilde 
\phi(y,\beta)$ it yields six Dirac-delta functions  which 
impose the following constraints on the momenta
\begin{eqnarray}
(q_i^0-k^0)= \sqrt{M_i^2+ (\vec{q}_i -\vec{k})^2} \\
(r_i^0+k^0)= \sqrt{\delta_i^2+ (\vec{r}_i +\vec{k})^2}
\end{eqnarray}
Four of the Dirac-delta functions will eliminate the  loop  momentum $k$ integration completely. 
One will be 
the overall pole condition on the total energy(eq.(\ref{tot_ener})) and the other Dirac-delta function 
restricts the 
external momenta. The loop  momentum $k$ integration also yields a self-consistency condition(Landau 
equation) namely the vanishing of the following four momenta
\begin{equation}
\sum_i \Big ( -a_i(q_i-k)+ b_i(r_i+k)\Big ) =0
\label{landau}
\end{equation}
for some $a_i, b_i \geq 0$. The physical interpretation of this equation is that if the masses of all 
lines in the graph of Fig[\ref{fig:proton}] are taken to be real, then the process depicted in this 
graph can occur 
in 
real space-time classically provided they obey this constraint\cite{coleman}. 
The total energy is 
\begin{equation}
E=\!\sum_{i=1}^{3}\!(q_i^0+r_i^0)\!=\!\sum_{i=1}^{3}\!\sqrt{M_i^2\!+\!(\vec{q}_i 
\!-\!\vec{k})^2}\!+\!\sqrt{\delta_i^2\!+\! 
(\vec{r}_i\!+\!\vec{k})^2}
\label{tot_ener}
\end{equation}
Minimising $E$, all spatial momenta vanish and we get the lowest baryon mass $M_B$, 
\begin{equation}
M_B=\sum_{i=1}^{3} ( M_i+\delta_i)
\end{equation}
For this case, the conditon in eq.(\ref{landau}) can be satisfied trivially.
 The mass of the lightest  baryon, the proton, is in terms of the mass of the pion taken to 
be $M_\pi = 139Mev$ and the size of the pion,$\delta_\pi$, inferred to be $174Mev(1fm)$. 

 The picture that we have deduced 
from this 
analysis is precisely that of the constituent quark model. The interpretation of the 
graph(Fig[\ref{fig:proton}]) is 
that there is a pole corresponding to a baryon state $|B\rangle$ and corresponds to the following 
matrix 
element,
\begin{equation}
\langle qqq|B\rangle \frac{1}{s-M_B^2+i\epsilon}\langle B|qqq \rangle
\end{equation}
where $s=E^2$ in the rest frame of the Baryon. Consequently we  infer the overlap function 
$\langle B|qqq \rangle$ in terms of the explicit meson wave-funcitons.
Furthermore for the lightest baryons the three quark momenta $p_i$ are such that they are 
constrained to
be $p_i^2=(313Mev)^2$. In the rest frame of the baryon these energies of the quarks   decompose
into 
masses $M_\pi$ of the mesons and $\delta_\pi$(bound quark mass in the meson) as shown by the cut in 
Fig[\ref{fig:proton}]. 

 In the theory Fig[\ref{fig:proton}] is not the only pole contribution to the six quark Green's 
function. Indeed 
there are many others. Firstly this 
graph 
can repeat several times by itself or repeat with other quark propagations in between. All these also 
have singular Dirac-delta function contributions corresponding to the baryon pole. We envisage a 
systematic expansion wherein these 
contributions result in a finite renormalisation of the baryon mass $M_B$. Therefore our estimate 
for
$\delta_\pi=174Mev$ subsumes these contributions. In 
't Hooft's $1 \over N$ expansion\cite{large_n1} where $N$ is the number of colors, we know 
that every additional 
meson 
function implies a factor of $1 \over \sqrt N$. Some of  these finite renormalisations are controlled 
in the $1\over N$ expansion. Indeed these contributions makes the evaluation of masses of  excited 
baryons tedious and also detailed  knowledge of meson wave-functions becomes necessary. 

\begin{figure}
\includegraphics[height=3cm]{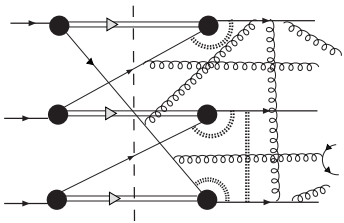}
\caption{Baryon pole(Parton picture).}
\label{fig:parton}
\end{figure}

 In addition there are other corrections which allows us to infer Feynman's parton model. For 
example,
using the same technique we can look at Green's functions where the three quark state goes to three 
quarks and several other gluons and quark anti-quark pairs as shown in Fig[\ref{fig:parton}]. 
All these graphs also have 
the same pole contribution as shown by the cut in Fig[\ref{fig:parton}] and indeed this graph 
corresponds to
\begin{equation}
\langle qqq|B \rangle \frac{1}{s-M_B^2 +i\epsilon} \langle B|qqq, gg (etc)\rangle
\end{equation}
where $g$ stands for gluons. From this we infer ~~~~$\langle B|qqq, gg (etc)\rangle $ matrix 
elements 
explicitly. That 
is 
the 
baryon  state has a finite probability amplitude to produce various other degrees of freedom in 
additon 
to the valence quarks.
\begin{equation}
\langle B|=\langle B|qqq \rangle \langle qqq|+\langle B|qqqgg\rangle \langle qqqgg| +\cdots
\label{bar_exp}
\end{equation}
From our analysis we see that each term on the the $r.h.s$ of the above equation has the same mass 
eigenvalue $M_B$ as dictated form Fig[\ref{fig:parton}].
It should be noticed that the constitutent quark picture of Fig[\ref{fig:proton}] is disturbed 
because these constituents can also exchange energy as shown in the Fig[\ref{fig:parton}] leading to
parton distribution functions. We 
also 
note in the 
conventional parton picture, quarks with momenta $\bar p_i$(Fig[\ref{fig:proton}])  emit gluons but 
we find in general 
there are quarks with lower energy, of the order of $\delta_i$ which  also emit gluons as shown in 
Fig[\ref{fig:parton}]. Although formally all these 
are precisely given in terms of meson wave-functions and quark and gluon propagators, a systematic 
convergent expansion is an open problem. 

  Stability of the baryon state would require that~~~ $M_B \leq 3\tilde m$. 
We have no {\it a priori} estimate of $\tilde m$ in QCD but a self-consistent picture to be presented 
later shows that $\tilde m \leq 591Mev$. In a model theory such as $\sigma$QCD\cite{ramesh} 
it has been estimated\cite{quark} for $u,d,s$ quarks to be about $550 Mev$. 

 Excited 
baryons can come about because of excited mesons in the intermediate states of Fig[\ref{fig:proton}]. 
Note that 
excited mesons such as $\rho,\omega$ $etc$ are necessarily unstable becuase they do decay into 
$2\pi$ 
or $3 \pi$ states. Consequently the meson pole is in the complex energy domain. Our analysis of the 
Green's functions can be carried out even in the complex plane consistent with causality ($i.e.,$ 
with Feynman $i\epsilon$ prescription) namely positive energy pole has negative imaginary part and 
{\it vice versa}.  The external momenta $p_i$ and $\bar{p}_i$ then have to  be  complex. As a 
consequence the width of the baryon is necessarily greater than  the widths of these 
intermediary mesons. For example if we consider intermediate mesons to be $\rho$ and $\pi$ 
with masses $M_\rho$ and $2 M_\pi$, we get $M_B=2(M_\pi 
+\delta_\pi)+ (M_\rho+\delta_\rho) \approx 1474Mev$ for $\delta_\rho \approx {\delta_\pi \over 
2}$(since the size of the $\rho$ meson has to be larger than that of the the pion). This matches with 
the 
Roper resonance and we would also find that the widths obey $\Gamma_{Roper} > \Gamma_\rho \approx 150 
Mev$, which is 
consistent with phenomenology. As we  see in this example predicting higher states demands 
knowing 
meson properties, $M_n$ and $\delta_n$. The latter being unavailable we have to develop theoretical 
models. 
As another example, consider the $\Lambda$ particle which is made up of $u,d,s$ quarks. The intermediate 
mesons in Fig[\ref{fig:proton}] have to be two $\pi$  and $K^0$. From the masses of these mesons we 
get the mass of 
the $\Lambda$ baryon, $M_\Lambda =1.2Gev$ for $\delta_K \approx \frac{\delta_\pi}{ 2}$. 

 Our method of estimating the spectrum of baryons although unconventional is consistent with 
quantum field 
theory. It is also consistent with the Hamiltonian eigenvalue problem. The complication is that the 
eigenvalue problem is stated in the basis of infinite Fock states. It is interesting to note  that 
the same eigenvalue problem
(eq.(\ref{bar_exp})) can be 
solved by using Green's functions and their singularities. We can explicitly get the mass eigenvalues 
and baryon states in terms of   meson states.  Now, it has to noted that if we take a system 
which is essentially non-relativistic, $i.e.,$ the threshold masses are large compared to the binding 
energy, $b_i$, then the mass of the 2-body meson bound state is given by, $M_i \approx 2 \tilde m 
-b_i$ and 
$\delta_i \approx a 
\tilde m$ where $a$ is some interaction coupling constant. Consequently we will estimate the baryon 
mass from Fig[\ref{fig:proton}] to be $3( 2 \tilde m +a \tilde m -b_i)$ which is much greater 
than $3 \tilde m$ 
(mass of three free quark states) implying that the system is unstable. This shows our procedure is 
inapplicable for all heavy quark systems. Similar considerations on charm and $u,d$ quark systems also 
results in instability for the baryon bound state. 
 
 Exact chiral symmetry is another limit where we can draw conclusions. This as discussed in 
\cite{quark} is a singular limit.
Due to Goldstone's theorem we have an exact relationship between the pion and the quark propagator 
\cite{jackiw} namely $\phi_\pi(p) \propto \gamma_5 Tr S(p)$ showing that  $\delta_\pi =\tilde m$ and 
also $M_\pi=0$. Consequently the mass of the baryon is $M_B=3 \tilde m$ implying that the baryon is 
almost 
unstable. 
Hence in the constituent 
quark model estimate, the baryon cannot be bound. However there are finite renormalisation effects 
alluded to earlier. These effects  can give rise to a weakly bound baryon 
state.  

\begin{figure}
\includegraphics[height=5cm]{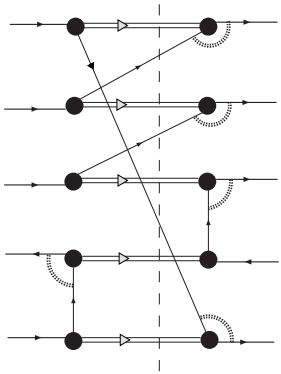}
\caption{Pentaquark pole.}
\label{fig:penta}
\end{figure}

Now consider the recently discovered\cite{penta} pentaquark candidate which has been discussed in the 
literature\cite{jaffe}. This is a $ud ud\bar 
s$ color singlet system with mass $1540 Mev$. In our picture the graph and cut shown in 
Fig[\ref{fig:penta}] gives 
the lowest pentaquark possible.
The mass of the pentaquark $\theta$ is therefore
\begin{equation}
M_\theta\!=\!3(M_\pi\!+\!\delta_\pi\!)\!+\!M_K\!+\!M_\pi\!=\!M_P\!+\!M_K\!+\!M_\pi\!=\!1571 Mev
\end{equation}
where $M_P$ is the mass of the proton or neutron and $M_K$ is the mass of the $K$ meson. Interestingly 
this 
prediction does not explicitly depend on $\delta_\pi$ if the mass of the proton is given. 
Fig[\ref{fig:penta}] shows 
that the spin of this state is $1\over 2$ as in the case of the proton.  As for as the 
width is concerned, since it is made up of only stable mesons, it is expected to be 
narrow. 

 We may look at other
pentaquarks such as $ud ud \bar u$, our prediction for its mass will be $M_P+ 2M_\pi =1217Mev$ and spin 
$1 \over 2$, which 
is 
close to the mass of the $\Delta$ baryon. This leads us to suspect that in the standard $\Delta$ 
resonance which has  a width of about $120Mev$ there exists a pentaquark resonance as well. The 
standard $\Delta$ 
resonance  is  the $uuu$ state which in our picture comes about from  symmetry 
considerations in the same way as in the constituent quark model. In terms of graphs, there are 
actually two graphs of the Fig[\ref{fig:proton}] type and the cancellations 
between them yields 
a higher mass to the $uuu$ system than an $uud$ system. 
 Similarly the masses of heptaquark states such as $ud us \bar d s \bar u$ can also be predicted to be
$M_P+ 2 M_K+2 M_\pi=2313 Mev$ and this resonance will be a little broader  since it has 
many decay channels. 

 In our considerations here, the important infrared properties of the quarks and mesons that are 
relevant to the description of baryons are the threshold masses, $\tilde m$, meson masses, $M_n$ 
and sizes, $\delta_n$. We have experimental knowledge of $M_n$ and to a lesser extent of $\delta_n$ 
inferred from form factors. In principle $\tilde 
m$ and $\delta_n$ can be inferred from Lattice simulation of QCD. Another attempt is to model QCD $q 
\bar q$ scattering kernel in the $BS$ equation(Fig[\ref{fig:scat}b])in a reliable manner. 
$\sigma 
$QCD\cite{ramesh} 
is  one such approach where the kernel is  taken to be purely a one gluon exchange and a singular 
$'{\sigma 
\over q^4}'$ propagator. Here $q$ is the exchanged momentum between the quark and the anti-quark. 
This was 
analysed 
in the large $N$ expansion with $g^2N$($g$ is the gauge coupling constant) also small. In this theory 
which is asymptotically free, we find spontaneous symmetry breakdown of chiral symmetry and 
PCAC\cite{quark,meson}. The 
threshold mass $\tilde m$ of $u,d,s$ quarks is numerically estimated to be $550Mev$ and $\delta_\pi$ 
can be estimated. 
 
 In the above discussion even 
if the 
quark is not confined $i.e.,$ the quark propagator has a pole at $\tilde m$, as long as the 
inequalities alluded to in the paper are satisfied the results remain true. Color non-singlets 
such as diquark 
systems can 
also be investigated in our picture. The graph of this system is that of Fig[\ref{fig:proton}] but 
with with the 
middle quark line removed. In $4$ dimensions the loop momentum is uniquely fixed and there is no 
residual Dirac-delta function to restrict the total energy of the Green's function. This implies the 
diquark system has no pole but only a branch cut 
singularity. This may have important consequences in  scattering processes. Similarly if we take 
a four quark 
system, they can be bound like the three quark system  with a mass $4(m_\pi + \delta_\pi)$ even 
though they can never be a color singlet.
Looking at non-singlets one can consider a $qq\bar q$ system which has the same color quantum numbers as 
a fundametal quark. This yields  the same graph as in Fig[\ref{fig:penta}] but without two 
quarks. The estimate of the mass of this state being $3 m_\pi +\delta_\pi=591Mev$ and is 
necessarily a branch cut. Consequently 
consistency of our picture demands that the threshold mass be,
\begin{equation}
\tilde m \leq 591Mev
\end{equation}
Color singlet criterion has played very little role in our analysis, if at all these considerations 
have to come from other theoretical aspects of QCD.  It is assumed that mesons are 
color singlets. It should be borne in mind that production of these multi-quark  states becomes harder 
because these have to obey the criteria of eq.(\ref{landau})  in 
the real phase space.

\end{document}